\documentstyle[11pt,aaspp4]{article}
\slugcomment{Submitted to the Astrophysical Journal}
\input psfig.sty

\def\w44{W\thinspace{44}}
\def\3c391{3C\thinspace{391}}
\def\w28{W\thinspace{28}}

\begin{document}

\title{OH(1720 MHz) Masers As Signposts of Molecular Shocks}

\author{Dale A. Frail\altaffilmark{1} and
George F. Mitchell\altaffilmark{2}}
 
\altaffiltext{1}{National Radio Astronomy Observatory, Socorro,
           NM, 87801, USA} 

\altaffiltext{2}{Department of Astronomy and Physics, Saint Mary's
  University, Halifax, NS B3H 3C3, CANADA}

\begin{abstract}
  We present observations of molecular gas made with the 15-m James
  Clark Maxwell Telescope toward the sites of OH(1720 MHz) masers in
  three supernova remnants: W\thinspace{28}, W\thinspace{44} and
  3C\thinspace{391}. Maps made in the $^{12}$CO $J=3-2$ line reveal
  that the OH masers are preferentially located along the edges of
  thin filaments or clumps of molecular gas. There is a strong
  correlation between the morphology of the molecular gas and the
  relativistic gas traced by synchrotron emission at centimeter
  wavelengths. Broad CO line widths ($\Delta$V=30-50 km s$^{-1}$) are
  seen along these gaseous ridges, while narrow lines are seen off the
  ridges. The ratio of H$_2$CO line strengths is used to determine
  temperatures in the broad-line gas of 80 K, and the $^{13}$CO
  $J=3-2$ column density suggests densities of 10$^4$-10$^5$
  cm$^{-3}$. These observations support the hypothesis that the
  OH(1720 MHz) masers originate in post-shock gas, heated by the
  passage of a supernova remnant shock through dense molecular gas.
  From the observational constraints on the density, velocity and
  magnetic field we examine the physical properties of the shock and
  discuss the shock-production of OH.  These OH(1720 MHz) masers are
  useful ``signposts'', which point to the most promising locations to
  study supernova remnant/molecular cloud interactions.
\end{abstract}


\vfill\eject
\section{Introduction}

It has recently been argued that a satellite line ($\nu_\circ=1720.53$
MHz) from the ground state of the hydroxyl radical (OH) is a powerful
tracer for studying the interaction of supernova remnant shocks with
molecular clouds (Frail, Goss \& Slysh 1993). Bright OH(1720 MHz)
maser emission has been detected towards 17 supernova remnants (10\%
of the observed sample) in our Galaxy (Green et al. 1997). On the
basis of this positional and velocity coincidence the case can be made
that the OH(1720 MHz) masers in supernova remnants are a separate
class of OH masers in our Galaxy, distinct from the OH masers in
star-forming regions and those in the circumstellar shells of
late-type stars.

Both observationally and theoretically there is support for the
hypothesis that the masers are collisionally excited by H$_2$ heated
by the passage of a shock.  Early theoretical work by Elitzur (1976)
showed that collisions of OH with H$_2$ can create a strong inversion
of the 1720 MHz line for a range of kinetic temperatures (25\ 
K$\leq$T$_k\leq200$\ K) and molecular gas densities (10$^3$
cm$^{-3}\leq{n_{\rm{H}_2}}\leq{10^5}$ cm$^{-3}$) which are typical of
the conditions expected in cooling post-shock clouds. Pavlakis \&
Kylafis (1996a,b) have included the effects of far infrared line
overlap (due to thermal and turbulent motions) and used newly computed
collisional cross sections between OH and H$_2$ to confirm Elitzur's
basic result for a limited range of T$_k$=100 to 200 K.

The observational evidence that the OH(1720 MHz) masers originate in
the post-shock gas is less compelling, relying mostly on indirect and
morphological indicators (see Green et al. 1997). What is missing for
the most part are clear kinematic and chemical signatures of a
molecular shock in the vicinity of the OH(1720 MHz) masers. This
situation has begun to change. For the well-studied IC\thinspace{433}
there is a clear kinematic shock signature at the location of the
masers (van Dishoeck, Jansen \& Phillips 1993), and for
W\thinspace{44} and 3C\thinspace{391} Reach and Rho (1996) detected
local emission maxima for the important cooling line of [OI] at the
location of several OH(1720 MHz) masers. In this paper we detail a
molecular line study undertaken toward the sites of several OH(1720
MHz) masers with the aim of testing whether they originate in
post-shock molecular gas.

\section{Observations}

The observations were made over 4 nights from 1997 July 20 to 23 using
the 15-m James Clark Maxwell Telescope (JCMT). We made $^{12}$CO
$J=3-2$ raster maps around four OH(1720 MHz) maser sites toward three
supernova remnants: W\thinspace{28}, W\thinspace{44} and
3C\thinspace{391}. The mapped regions in W\thinspace{28} and
W\thinspace{44} are designated as W\thinspace{28F}, W\thinspace{44E}
and W\thinspace{44F} in Claussen et al. (1997), while for
3C\thinspace{391} we mapped the region around the southernmost maser
(Frail et al. 1996). Details are given in Table 1. 

At several of the intensity peaks found in the CO maps, we obtained
spectra of a number of other molecules. These observations and results
will be discussed in detail in a subsequent paper. Here we make use
only of the $^{13}$CO $J=3-2$ (330.588 GHz), H$_2$CO 3$_{03}$-2$_{02}$
(218.222 GHz), H$_2$CO 3$_{22}$-2$_{21}$ (218.475 GHz), and H$_2$CO
5$_{05}$-4$_{04}$ (362.736 GHz) transiions.  The $3-2$ transitions of
H$_2$CO required the facility A-band receiver, with a beamwidth of
20\arcsec\ and a main beam efficiency of 0.79.

\section{Results}

In Plates 1 and 2 we show maps of integrated CO $J=3-2$ emission
together with representative spectra at several locations. The maser
positions are also superimposed on these maps. The velocity data cubes
of these pointings show a rich variety of complex structure, but we
will defer a detailed discussion of these for another paper. Here we
present evidence which indicates the existence of molecular shocks in
the vicinity of the OH(1720 MHz) masers.

It is apparent in Plates 1 and 2 that the masers are not located at
random with respect to the molecular gas but rather are found near the
peaks in the integrated CO $J=3-2$ maps.  In one striking example the
line of nine masers in W\thinspace{44E} nicely delineates the forward
edge of the molecular gas, closest to the shock. Claussen et al.
(1997) have suggested that this line of masers traces the shock front
from W\thinspace{44} which is moving eastward into the ambient
molecular gas. The morphology of this gas is also noteworthy as it is
concentrated in well-defined filaments or clumps.  These features are
largely unresolved by our 13$^{\prime\prime}$ beam in one dimension
but are 1 pc to 2 pc long in the other. In W\thinspace{44F}, for
example, a thin filament of gas can be traced from the northwest to
the southeast across the entire
70$^{\prime\prime}\times75^{\prime\prime}$ map. It is likely that the
true extent of these features is not revealed by these figures since
we have mapped only a small fraction ($<1\%$) of each supernova
remnant. With few exceptions, the masers are preferentially located
nearer to the edge of the shock (as traced by the non-thermal
emission) than the peak of the CO itself. This suggests that some
special conditions are required in order to produce a strong inversion
of the OH molecule.

Another interesting correlation can be discerned between the
integrated CO $J=3-2$ maps and the radio continuum images found in
Claussen et al. (1997) and Wilner et al. (1998). Near W\thinspace{44F}
the radio continuum contours have the same orientation as the
molecular filament and a local maxima in the non-thermal radio
continuum of W\thinspace{44F} coincides with the peak in the CO
$J=3-2$ map. Likewise, for 3C\thinspace{391} the radio continuum and
the molecular clump have the same curvature while for W\thinspace{28F}
a prominent "kink" can be seen in in both the molecular gas and the
radio continuum. Local increases in the synchrotron emissivity do not
require new particle acceleration, but rather can originate from the
compression of the magnetic field and existing relativistic electrons
in a radiative shock (Blandford \& Cowie 1982). While good
correlations have been noted between the optical line and radio
continuum emission of several supernova remnants (e.g. Cygnus Loop:
Straka et al. 1986), this is the first clear signature to our
knowledge of a synchrotron/molecular correlation.

Although the morphology of the molecular gas is suggestive of a shock
interaction, it is the velocity extent of the lines that are the most
telling. Spectra taken near the peaks and along the ridges of the
integrated CO $J=3-2$ emission (see Figs. 1 and 2) consistently show
broad, assymetric lines whose widths ($\Delta$V=30-50 km s$^{-1}$),
reminiscent of those seen toward IC\thinspace{433} (van Dishoeck et
al. 1993). Away from the bright peaks the line widths narrow
considerably ($\Delta$V$\leq$10 km s$^{-1}$) (e.g. Fig. 1). Moreover,
whenever it is possible to identify this less disturbed gas, its LSR
velocity agrees with that of any nearby masers to within a few km
s$^{-1}$.  Examples include 3C\thinspace{391} where a narrow line
component in CO emission at +105.5 km s$^{-1}$ is found widely
distributed in the region which includes the OH(1720 MHz) maser at
104.9 km s$^{-1}$ (see also Reach and Rho 1998). In W\thinspace{44E}
and W\thinspace{44F} there are what appear to be narrow,
self-absorption features at +43 and +46 km s$^{-1}$, respectively.
Although they too are close to the mean LSR velocity of the masers in
each direction, some caution is warranted since false absorption may
be introduced by our sky switching calibration cycle.  The
correspondence between the maser velocities and colder, presumably
unshocked ambient medium is expected since it has been argued
elsewhere (Claussen et al. 1997) that these masers originate in shocks
which are viewed largely transverse to the line of sight.

For the bright CO peak at (15\arcsec,$-$40\arcsec) in the
W\thinspace{28F} map (Plate 1), we have also observed the three
H$_2$CO transitions 3$_{03}$-2$_{02}$, 3$_{22}$-2$_{21}$, and
5$_{05}$-4$_{04}$ to determine both gas kinetic temperature and gas
density.  The importance of H$_2$CO in measuring T$_k$ and n is
described in detail in Mangum \& Wootten (1993). The intensity ratio
3$_{03}$-3$_{02}$/3$_{22}$-3$_{21}$ provides the gas kinetic
temperature, either from LVG analysis (e.g. Mangum \& Wootten 1993;
van Dishoeck et al.  1993) or using LTE expressions (Mangum \& Wootten
1993). The observed intensity ratio in W\thinspace{28F} is 3.25,
implying T$_k$=80$\pm$10 K. This high value of T$_k$ is further
evidence of a shock. When T$_k$ is known, the ratio
3$_{03}$-2$_{02}$/5$_{05}$-4$_{04}$ provides the gas density.  With
T$_k$=80 K, we interpolated between LVG models of Mangum \& Wootten
(1993), using the observed 3-2/5-4 ratio of 1.5 to find
$n_{\rm{H}_2}=2\times{10}^{6}$ cm$^{-3}$.

We have $^{13}$CO $J=3-2$ spectra at $^{12}$CO intensity peaks in
W\thinspace{28F}, W\thinspace{44E}, and 3C\thinspace{391}. Because the
critical density for excitation of CO is low, level populations
probably follow a Boltzmann distribution.  Assuming LTE, it is
straightforward to obtain from the line ratio $^{12}$CO/$^{13}$CO the
optical depth, the excitation temperature, and the CO column density
(e.g. Mitchell et al. 1992). The observed beam-averaged column
densities yield a gas number density if we know the line-of-sight
extent of the gas. On the assumption that the gas in the beam has an
extent of one beamwidth, typical gas densities of
$n_{\rm{H}_2}={10}^{4}-{10}^{5}$ cm$^{-3}$ are found.


\section{Discussion}

These JCMT observations have demonstrated that the OH(1720 MHz) masers
in W\thinspace{44E}, W\thinspace{44F}, W\thinspace{28F} and
3C\thinspace{391} are located near local peaks in the integrated CO
$J=3-2$ emission. This hot, dense molecular gas is organized into thin
filaments or clumps, with the long axis perpendicular to the shock
normal (as inferred from the radio continuum), suggesting compression
of the gas by a passing shock. We deduce the existence of the shock
from the large observed velocity extent ($\Delta$V=30-50 km s$^{-1}$) and
the high measured temperature of 80 K. These observations lend
important support to the hypothesis that the OH(1720 MHz) masers
originate in the molecular gas behind the shock. We will now attempt
to infer the physical properties of the post-shock gas, determine the
character of the shock, and investigate models for the shock
production and excitation of OH.

In a radiative shock the shock velocity V$_s$ is small enough or the
ambient gas density $\rho_\circ$ is large enough that the timescale
for the gas to radiate away its thermal energy, acquired by the
passage of the shock, is short compared to the dynamical time of the
shock itself (Draine \& McKee 1993). The amount of compression
expected in this post-shock gas depends in large measure on the
strength of the initial magnetic field B$_\circ$. The magnetic
pressure dominates over the thermal pressure and therefore
$\rho_{ps}/\rho_\circ$=$\sqrt{2}\thinspace$V$_s$/V$_{\rm{A}}$, where
V$_{\rm{A}}$ is the Alfv\'en velocity in the ambient gas. It has been
customary to estimate B$_\circ$ using an empirical relation between B
and number density $n$ (e.g. Fiebig \& G\"usten 1989), which shows
that over at least 8 orders of magnitude in density
B($\mu{\rm{G}})=\sqrt{n\thinspace({\rm{cm}}^{-3})}$ and implies that
V$_{\rm{A}}\simeq$1.84 km s$^{-1}$. Such a relation is not unexpected
if there is rough equilibrium between the kinetic and magnetic energy
densities in molecular clouds (Myers \& Goodman 1988). However, the
large observed scatter makes it unreliable to use in individual cases.

The OH(1720 MHz) maser line allows a unique measurement to be made of
the strength of the line-of-sight magnetic field B$_{los}$ in the
post-shock gas using the Zeeman effect (Claussen et al. 1997). In
W\thinspace{44} and W\thinspace{28} Claussen et al. (1997) derived an
average B$_{los}$=200 $\mu$G which was remarkably constant in both
direction and strength across each remnant.  The large scale
distribution of magnetic field vectors toward W\thinspace{44} (Kundu
\& Velusamy 1972) as traced by the radio synchrotron emission is also
uniform. This suggests that these B$_{los}$ values are not some
peculiar local values, valid only in the vicinity of the masers, but
rather they represent some global measurement in the post-shock
magnetic field. For a randomly oriented field, the median value of the
total magnetic field strength is 2$\times$B$_{los}$ and therefore in
W\thinspace{44} and W\thinspace{28} the post-shock magnetic field
B$_{ps}$=400 $\mu$G, and the magnetic pressure
B$_{ps}^2/8\pi=6.4\times{10}^{-9}$ erg cm$^{-3}$, or alternatively
(see below) $2.7\times{10}^{5}$ cm$^{-3}$ (km s$^{-1}$)$^2$.


The magnetic pressure dominates over all other sources of pressure in
the post-shock gas, balancing the ram pressure of the gas entering the
shock wave $\rho_\circ$V$_s^2$. Thus a measurement of B$_{los}$ fixes
a line in the $n_\circ$-V$_s$ plane which constrains the properties of
the post-shock gas from which the OH(1720 MHz) masers originate. In
practice there is sufficient uncertainty in B$_{ps}$ that we will
consider shocks with a range of ram pressures over 10$^5$ to 10$^6$
cm$^{-3}$ (km s$^{-1}$)$^2$. Broadly speaking there are two classes of
solutions allowed, a dissociative J-shock and a non-dissociative
C-shock (Draine \& McKee 1993). In a J-shock (for V$_s>$25-45 km
s$^{-1}$) the physical conditions change abruptly across the shock and
molecules can be destroyed. In a C-shock the ions, drifting ahead of
the neutrals, produce a more gradual transition and a thicker heating
region with limited dissociation of molecules.

If we make the standard assumption that the observed CO $J=3-2$ line
widths $\Delta$V of 30-50 km s$^{-1}$ are giving the shock velocity
V$_s$, then from our ram pressure constraint we infer
$n_\circ\simeq{10}^2-{10}^3$ cm$^{-3}$. The final compression in the
post-shock gas will depend on V$_{\rm{A}}$ in the ambient gas but for
the value given above $n_{ps}\simeq{10}^4$ cm$^{-3}$. At these
relatively low densities the shock is likely to be a dissociative
J-type (Draine et al. 1983).  A significant column of OH can be formed
in the post-shock gas in such a shock (Neufeld \& Dalgarno 1989) but
the final column density of H$_2$ never gets sufficiently high to
prevent the eventual destruction of the OH by UV photons from the
leading edge of the shock. This might explain why the OH masers appear
to be confined along the leading edge of the integrated CO maps (see
\S{3}). There may be a small region behind the shock where the density
and temperature of the OH is sufficient to produce OH(1720 MHz) maser
emission, while further downstream the OH is photodissociated.


Another alternative that satisfies our ram pressure constraint would
be a slow, non-dissociative C-type shock propagating into denser gas.
The true V$_s$ could be smaller than the observed linewidths if the
13\arcsec\ beam intersected several shocks along the line-of-sight.
Furthermore, the detection of H$_2$CO at the peaks of the integrated
CO maps requires the presence of high density post-shock gas
$n_{ps}\simeq{10}^5-10^6$ cm$^{-3}$.  Theoretical models (Draine,
Roberge \& Dalgarno 1983, Kaufman \& Neufeld 1996) show that
significant columns of warm (T$_k>$1000 K) OH are produced for
V$_s>$15 km s$^{-1}$ at a few$\times{10}^{16}$ cm behind the leading
edge of the disturbance. However, as the gas cools to T$_k\sim$400 K
the models predict that virtually all of the atomic oxygen not tied up
in CO will be rapidly converted to H$_2$O.  Wardle, Yusef-Zadeh \&
Geballe (1998) suggested a novel solution to this problem,
dissociating the H$_2$O by the soft X-rays emitted by the hot gas
interior to the remnant. The X-rays penetrate into the dense
water-rich gas producing OH in a small region ($\sim{10}^{15}$ cm)
over which T$_k\sim$100-200 K.  We consider the Wardle et al. (1998)
model to be a particularly promising one since it would explain the
apparent narrow region over which we find the OH masers (with respect
to the molecular gas) and our measured T$_k$ value.

\acknowledgments

We thank Bill Reach and Jeonghee Rho for communicating their IRAM
results prior to publication and we thank Lorne Avery and Gerald
Moriarty-Schieven for their help in setting up the JCMT observing.
GFM is supported by the Natural Sciences and Engineering Research
Council of Canada. DAF is supported by the NRAO. The NRAO is a
facility of the National Science Foundation operated under cooperative
agreement by Associated Universities, Inc. The James Clerk Maxwell
Telescope is operated by The Joint Astronomy Centre on behalf of the
Particle Physics and Astronomy Research Council of the United Kingdom,
the Netherlands Organization for Scientific Research, and the National
Research Council of Canada.

\clearpage

\begin{table}
\small
\begin{flushleft}
\begin{tabular}{cccccc}
\multicolumn{6}{c}{\sc Table~1 -- Telescope
  Pointing Parameters for CO $J=3-2$ Mapping}
\\
\hline
\hline
Name & RA(B1950) & Dec(B1950) & V$_{\rm{LSR}}$ & Ref. & Raster \\
     & (h m s) & ($\arcdeg$ $\arcmin$ $\arcsec$) & km s$^{-1}$ &
     (\arcsec,\arcsec) &(\arcsec$\times$\arcsec) \\
\hline
W\thinspace{28F}  & 17 58 49.2  & $-$23 19 00.0 & 12  & 0,$-$1200      & 75$\times$110 \\
3C\thinspace{391} & 18 46 47.7  & $-$01 01 00.0 & 105 & $-$120,120     & 80$\times$80 \\
W\thinspace{44E}  & 18 53 57.0  & $+$01 25 45.0 & 45  & $-$760,$-$1440 & 75$\times$120 \\
W\thinspace{44F}  & 18 54 04.7  & $+$01 22 35.0 & 45  & $-$760,$-$1440 & 70$\times$105 \\
\hline
\hline
\tablecomments{(1) The facility SIS receiver (B3) was used at a center
  frequency of 345.796 GHz and the signal was fed into a digital
  autocorrelation spectrometer for a bandwidth of 500 MHz and channel
  spacing of 378 kHz (0.3 km s$^{-1}$).  (2) The raster maps were made
  by driving the telescope in right ascension and integrating for 3
  seconds at each position separated by 5$^{\prime\prime}$. (3) Sky
  subtraction was effected by observing a reference position after
  each row. (4) The beamsize at this frequency is 13$^{\prime\prime}$
  and the main beam efficiency is 0.62.}
\end{tabular}
\end{flushleft}
\dummytable\label{line-fits-table} 
\end{table}
\clearpage

\medskip
\noindent {\bf Plate  1.--} (left) A 75\arcsec$\times$110\arcsec\ map of
$^{12}$CO $J=3-2$ towards a region of OH(1720 MHz) masers known as
W\thinspace{28F}. The masers locations are indicated by the filled red
circles. The spectra are integrated over the bulk of the emission from
$-$20 km s$^{-1}$ to +40 km s$^{-1}$. The color wedge on the right
hand side of the plot is the velocity-integrated antenna temperature
in units of K km\ s$^{-1}$. The shock front, determined from the
centimeter radio continuum, lies toward the east. (right) Similar to
W\thinspace{28F} but of a 75\arcsec$\times$120\arcsec\ region toward
W\thinspace{44E}. The spectra are integrated from +20 km s$^{-1}$ to
+38 km s$^{-1}$. The shock front lies toward the north east. At the
assumed distance of 3 kpc for both W\thinspace{28} and W\thinspace{44}
the 13\arcsec\ beam has a spatial resolution of 0.2 pc.

\noindent{Copies of this plate can be found at 
http://www.nrao.edu/$\sim$dfrail/jcmt\_plate1.gif}

\clearpage
\medskip
\noindent {\bf Plate 2.--} (top) Similar to Plate 1 but of a 
80\arcsec$\times$80\arcsec\ region toward 3C\thinspace{391}. The
spectra are integrated from +60 km s$^{-1}$ to +150 km s$^{-1}$. The
shock front lies toward the south west. At the assumed distance of 7.2
kpc for 3C\thinspace{391} the 13\arcsec\ beam has a spatial resolution
of 0.45 pc. (Bottom) Similar to 3C\thinspace{391} but of a
75\arcsec$\times$105\arcsec\ region toward W\thinspace{44F}.  The
spectra are integrated from +49 km s$^{-1}$ to +60 km s$^{-1}$.  The
shock front lies toward the north east.

\noindent {Copies of this plate can be found at 
http://www.nrao.edu/$\sim$dfrail/jcmt\_plate2.gif}

\clearpage
\centerline{\hbox{\psfig{file=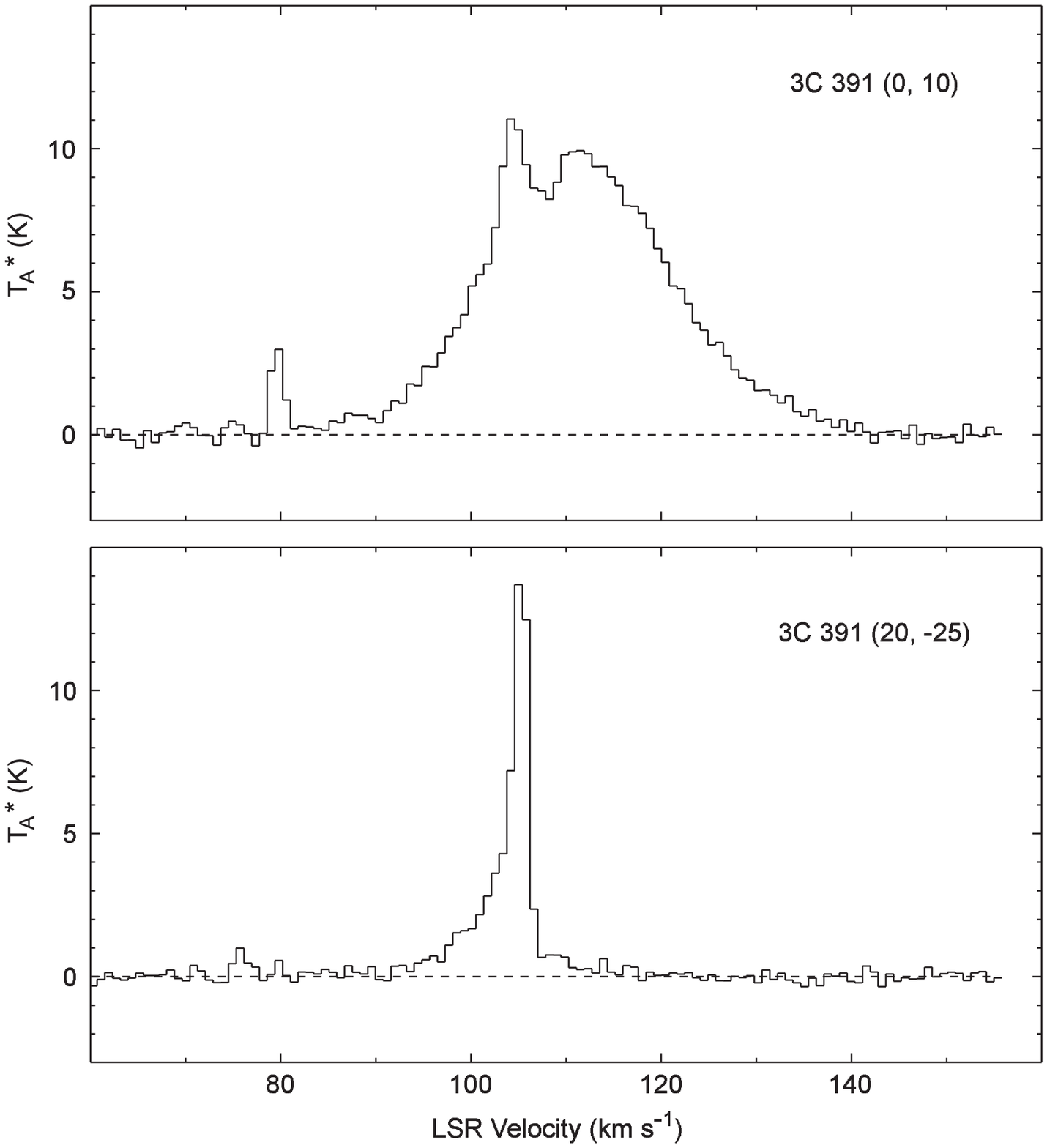,width=5.0in,clip=0,angle=0}}}
\medskip
\noindent {\bf Figure 1.--} Sample spectra taken from the $^{12}$CO $J=3-2$
map of 3C\thinspace{391} (Plate 2) at the peaks of the narrow line and
broad line emission areas (see main text). These peaks are offset from
the map center by (20\arcsec,$-$25\arcsec) and (0\arcsec,10\arcsec),
respctively. The OH(1720 MHz) masers velocity is +104.9 km s$^{-1}$.
The line brightness on the vertical axis is expressed as antenna
temperature T$_{\rm{A}}^*$.

\clearpage
\centerline{\hbox{\psfig{file=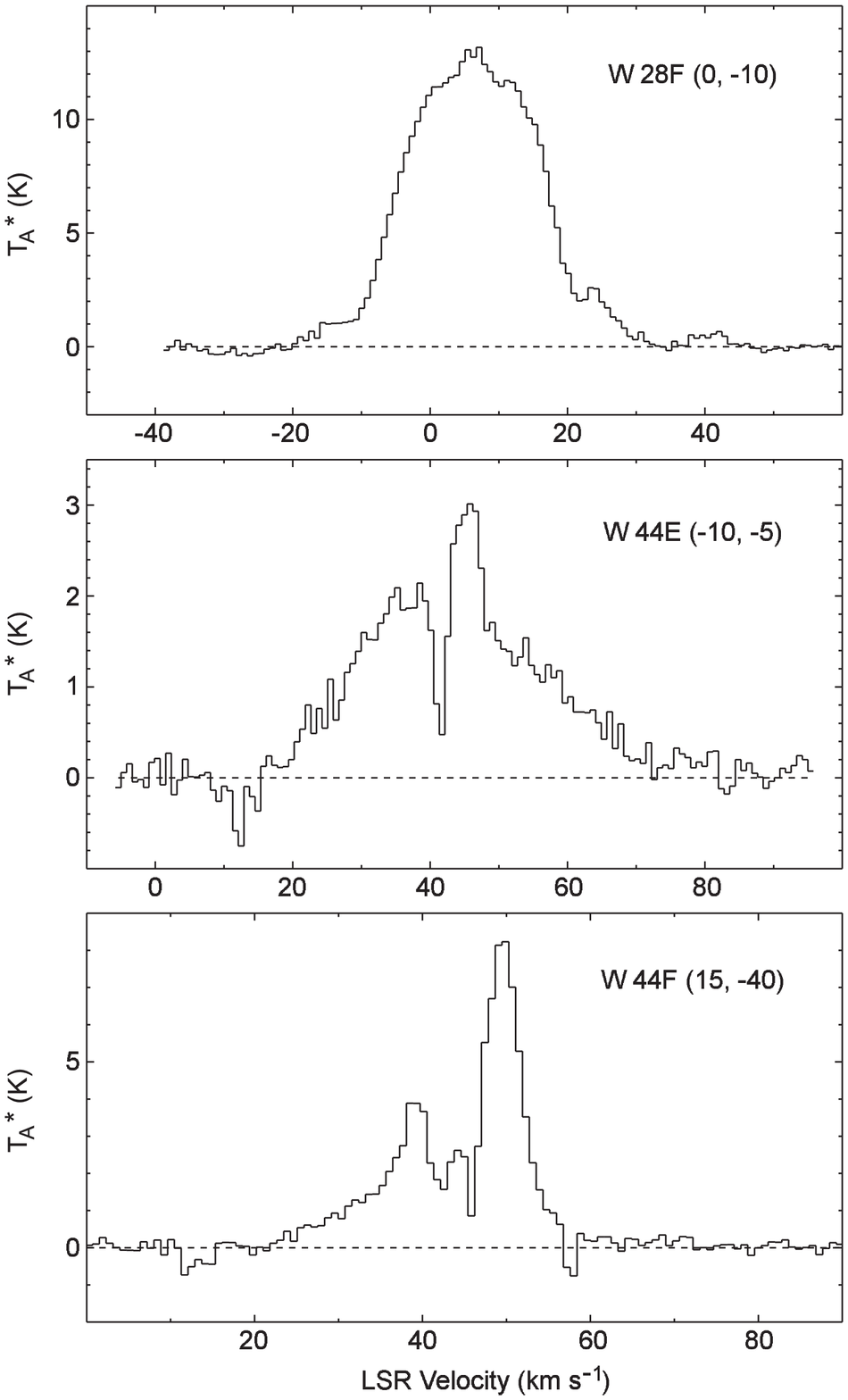,width=5.0in,clip=0,angle=0}}}
\medskip
\noindent {\bf Figure 2.--} Sample spectra from the integrated 
$^{12}$CO $J=3-2$ maps of W\thinspace{28F}, W\thinspace{44E} and
W\thinspace{44F}. The spectra are taken towards the peaks at position
offsets from the center of each map of (0\arcsec,$-$10\arcsec),
($-$10\arcsec,$-$5\arcsec) and (15\arcsec,$-$40\arcsec), respectively.
The pseudo absorption features are likely artifacts produced by
emission in the reference position. The average OH(1720 MHz) maser
velocities for W\thinspace{28F}, W\thinspace{44E} and W\thinspace{44F}
are +11.0 km s$^{-1}$, +44.7 km s$^{-1}$, and +46.6 km s$^{-1}$,
respectively.


\clearpage
 
\begin{references}

\reference {bc82} Blandford, R. D., \& Cowie, L. L. 1982, \apj, 260, 625


\reference {Cl97} Claussen, M. J., Frail, D. A., Goss, W. M., \&
Gaume, R. A. 1997, \apj, 489, 143

\reference {drd83} Draine, B. T., Roberge, W. G., \& Dalgarno,
A. 1983, \apj, 264, 485

\reference {dm93} Draine, B. T., \& McKee, C. F. 1993, \araa, 31, 373

\reference {e76} Elitzur, M. 1976, \apj, 203, 124


\reference {ehm89} Elitzur, M., Hollenbach, D. \& McKee, C. F. 1989, 
\apj, 346, 983

\reference {fg89} Fiebig, D., \& G\"usten, R. 1989, \aap, 214, 333

\reference {fgs93} Frail, D. A., Goss, W. M., \& Slysh, V. I.  1994,
ApJ, 424, L111

\reference {f96} Frail, D. A., Goss, W. M., Reynoso, E. M., Giacani,
E. B., \& Green, A. J. 1996, \aj, 111, 1651

\reference {ag97} Green, A. J., Frail, D. A., Goss, W. M., \&
Otrupcek, R. 1997, \aj, 114, 2058



\reference {kn} Kaufman, M. J., \& Neufeld, D. A. 1996, \apj, 456, 611

\reference {kv72}  Kundu, M. R., \& Velusamy, T. 1972, \aap, 20, 237


\reference {mw93} Mangum, J. G. \& Wootten, A. 1993, \apjs, 89, 123

\reference {mhs92} Mitchell, G. F., Hasegawa, T. I., \& Schella, J.
1992, \apj, 386, 604


\reference {mg88} Myers, P. C., \& Goodman, A. A. 1988, \apj, 326, L27

\reference {nd89} Neufeld, D. A., \& Dalgarno, A. 1989, \apj, 340, 869


\reference {pk96a} Pavlakis, K. G., \& Kylafis, N. D. 1996a, \apj, 467, 300

\reference {pk96b} Pavlakis, K. G., \& Kylafis, N. D. 1996b, \apj, 467, 309

\reference {rk96} Reach, W. T., \& Rho, J. -H. 1996, \aap, 315, L277

\reference {rk96} Reach, W. T., \& Rho, J. -H. 1998, in preparation



 
\reference {str86} Straka, W. C., Dickel, J. R., Blair, W. P., \&
Fesen, R. A. 1986, \apj, 306, 266

\reference {van93} van Dishoeck, E. F., Jansen, D. J., 
\& Phillips, T. G. 1993, \aap, 279, 541



\reference {wyg98} Wardle, M., Yusef-Zadeh, F., \& Geballe,
T. R. 1998, \mnras, submitted

\reference {wil98} Wilner, D. J., Reynolds, \& S. P., Moffett, D. A.
1998, \aj, 115, 247

\end{references}
\end{document}